\documentclass[a4paper,11pt]{article}
\usepackage{jinstpub} % for details on the use of the package, please
\usepackage{lineno}
\title{\boldmath Light outputs of yttrium doped BaF$_2$ crystals irradiated with neutrons}

\author{V. Baranov,}
\author[1]{Yu.I. Davydov\note{Corresponding author.}}
\author{and I.I. Vasilyev}

\affiliation{JINR, Dubna, Russia}

\emailAdd{davydov@jinr.ru}

\abstract{The fast luminescence component of barium fluoride (BaF$_2$) crystals with a subnanosecond decay time can find wide application in particle physics and nuclear physics. However, the slow luminescence component with the 630~ns decay time could cause pile-up signals at a high rate environment. Doping of BaF$_2$ crystals with rare earth elements suppresses the slow emission component, but at the same time the radiation hardness of the crystals deteriorates.
This work presents the results of studying crystal samples, both pure BaF$_2$ and those doped with yttrium in a proportion of 1at.\%Y, 3at.\%Y and 5at.\%Y, irradiated with a fast neutron fluence of about 2.3$\times$10$^{14}$~n/cm$^2$. Their light output and decay kinetics were measured before and after irradiation. It is found that the light output loss of a pure BaF$_2$ crystal after irradiation is about 7\%, and the light output loss of yttrium doped samples after irradiation is about two times higher. The measurement results demonstrate that after irradiation the fast component of each sample has a relative light output loss 2-3\% larger than the slow one.
}


\begin{document}
\maketitle
\flushbottom

\section{Introduction}
\label{sec:intro}

Barium fluoride (BaF$_2$) crystals have long been known as scintillators with a broad luminescence band with a peak at 310 nm and a decay time of about 630 ns.
Interest in studies of barium fluoride crystals increased after luminescence was discovered in them in the early 1980s at the boundary of the ultraviolet and vacuum ultraviolet regions (190-225 nm) with a decay time of about 0.7-0.8~ns~\cite{Ershov, Gudovskikh, Laval}.
The fast luminescence component of BaF$_2$ allows it to be used as a fast scintillator in medical applications, in high energy physics and nuclear physics.

The Mu2e experiment, which is currently being prepared at Fermilab, uses CsI crystals in the electromagnetic calorimeter \cite{Mu2e_EMC}. At the second phase of the experiment, it is planned to significantly increase the beam intensity, to use an electromagnetic calorimeter made of BaF$_2$ crystals to increase the speed of operation~\cite{Mu2e-II}. However, a high fraction of the slow component (about 85\%) in the total luminescence can cause pile-up of signals at a high beam rate. To solve this problem, various approaches are being investigated: 1) the use of thin films on an avalanche photodetector \cite{Hitlin} to suppress luminescence above 280~nm, 2) the use of a thin multilayer filter on the quartz glass substrate between the crystal and the photodetector to suppress luminescence in the range 250-400~nm ~\cite{Artikov_filter}, 3) the use of a solar-blind photomultiplier with an aluminum-gallium nitride (AlGaN) photocathode~\cite{Atanov_cathode}, 4) suppression of the slow scintillation component by doping a BaF$_2$ crystal with yttrium~\cite{Chen}.

It is known that doping of a BaF$_2$ with rare earth elements (La, Y, Lu, Sc) suppresses the slow component emission~\cite{Schotanus}. However, doping of crystals usually leads to a decrease in their radiation hardness~\cite{Woody}. In this paper, we present a comparison of the light outputs of pure and yttrium doped BaF$_2$ crystal samples before and after irradiation with a neutron beam.

\section{Crystal samples}
\label{sec:crystals}

In total, four samples were selected for the study: one pure BaF$_2$ crystal and three samples doped with a rare earth element yttrium in a proportion of 1at.\%Y, 3at.\%Y and 5at.\%Y. The samples were grown at SICCAS (China) by the Bridgman method. The samples were cut from large ingots and had a size of 1$\times$1$\times$1~cm$^3$, all faces were optically polished. Light outputs and decay kinetics of crystal samples were measured before and after irradiation.

\section{Irradiation of crystals}
\label{sec:irradiation}

The crystal samples were irradiated in channel \#3 of the IBR-2M~\cite{IBR-2M} pulsed reactor at the Frank Laboratory of Neutron Physics, JINR. The reactor operates with a pulse frequency of 5 or 10 Hz and a pulse duration of 200-300~$\mu$s. Channel \#3 was specifically built to study the effect of neutron and gamma radiation on materials.  The neutron flux density immediately after the water moderator is $\sim$10$^{16}$ n/(cm$^2\times$s) per pulse or $\sim$10$^{13}$ n/(cm$^2 \times$s) on average over time~\cite{IBR_Ch3}.
All four crystal samples were placed together about 5~m from the water moderator and received the same radiation dose. A nickel wire placed with the samples was used to measure the fast neutrons (E>1~MeV) fluence from the wire induced activity.

Irradiation of the samples was carried out during the working cycle of the reactor, which lasted 14 days. From the activity induced on the nickel wire it was found that about 2.3$\times$10$^{14}$~n/cm$^2$ (E>1~MeV) passed through the samples during the irradiation run.

\section{Results and discussion}
\label{sec:results}

The light outputs of the crystal samples were measured before and after neutron irradiation with a $^{22}$Na gamma source at room temperature. Two back-to-back outgoing 511 keV gammas from the $^{22}$Na source hit the crystal under tests and an additional trigger crystal of the same size and generated coincidence triggers. The light output of the samples was estimated from the full absorption peaks of 511~keV gammas. The systematic errors in measuring the positions of the full absorption peaks  did not exceed 1.5\%.

The samples were wrapped with a double layer of 0.1~mm thick Teflon film.  The signals from the BaF$_2$ samples were registered by a Hamamatsu R2059 photomultiplier tube (PMT). The measuring channel was calibrated using a single electron peak of the PMT. No optical grease was used between the samples and the PMT photocathode.  Signals from the R2059 PMT were recorded by the 10 bit 2~Gs/s CAEN NDT5751 Digitizer. The total signals from the BaF$_2$ samples were integrated within 2~$\mu$s. The fast luminescence component of the samples  was measured during the first 20~ns and the slow one after 20~ns.

The R2059 PMT has a quantum efficiency of about 16-17\% in the fast luminescence region (200-220~nm) and about 23-24\% in the slow luminescence region (310~nm). The measured light outputs of the fast and slow luminescence of the samples were not corrected for the difference in the quantum efficiency of the PMT.

\begin{figure}[!ht]
  \begin{center}
   \includegraphics[width=0.32\textwidth]{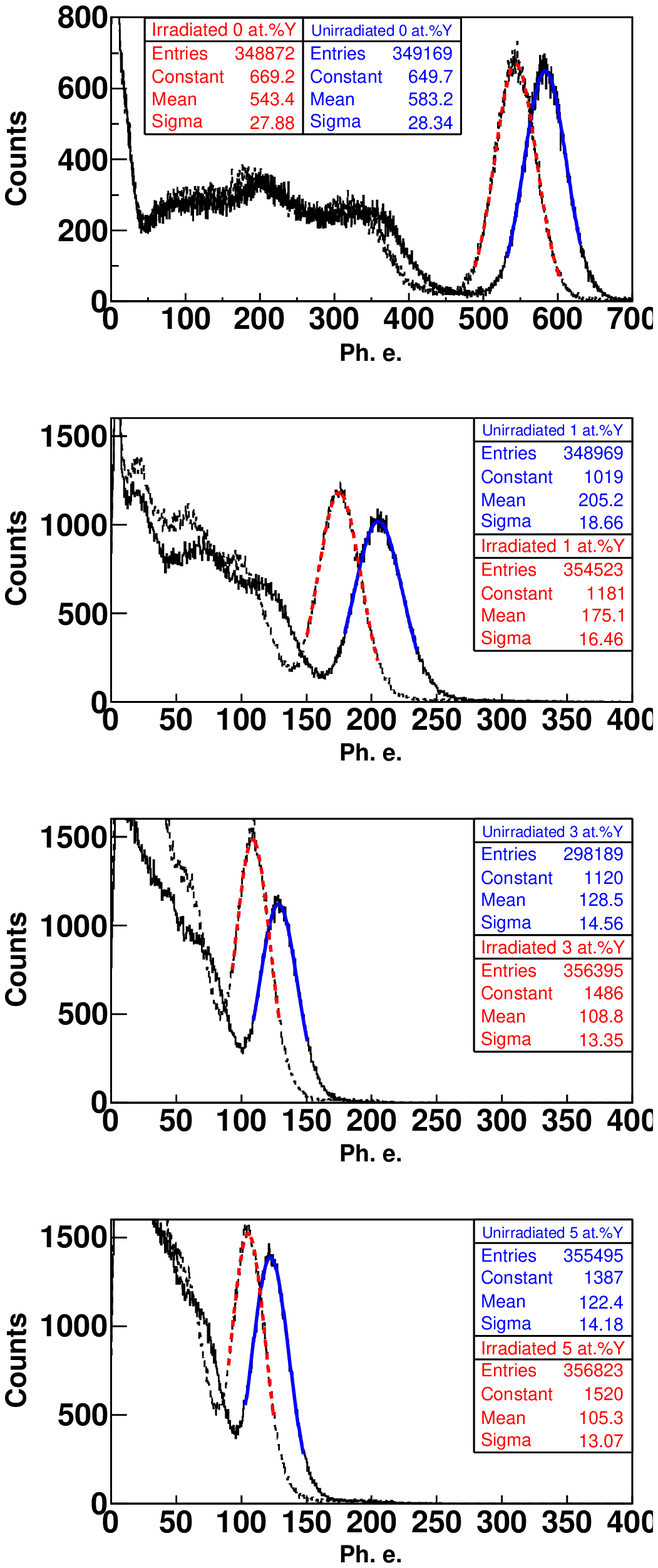}
    \includegraphics[width=0.32\textwidth]{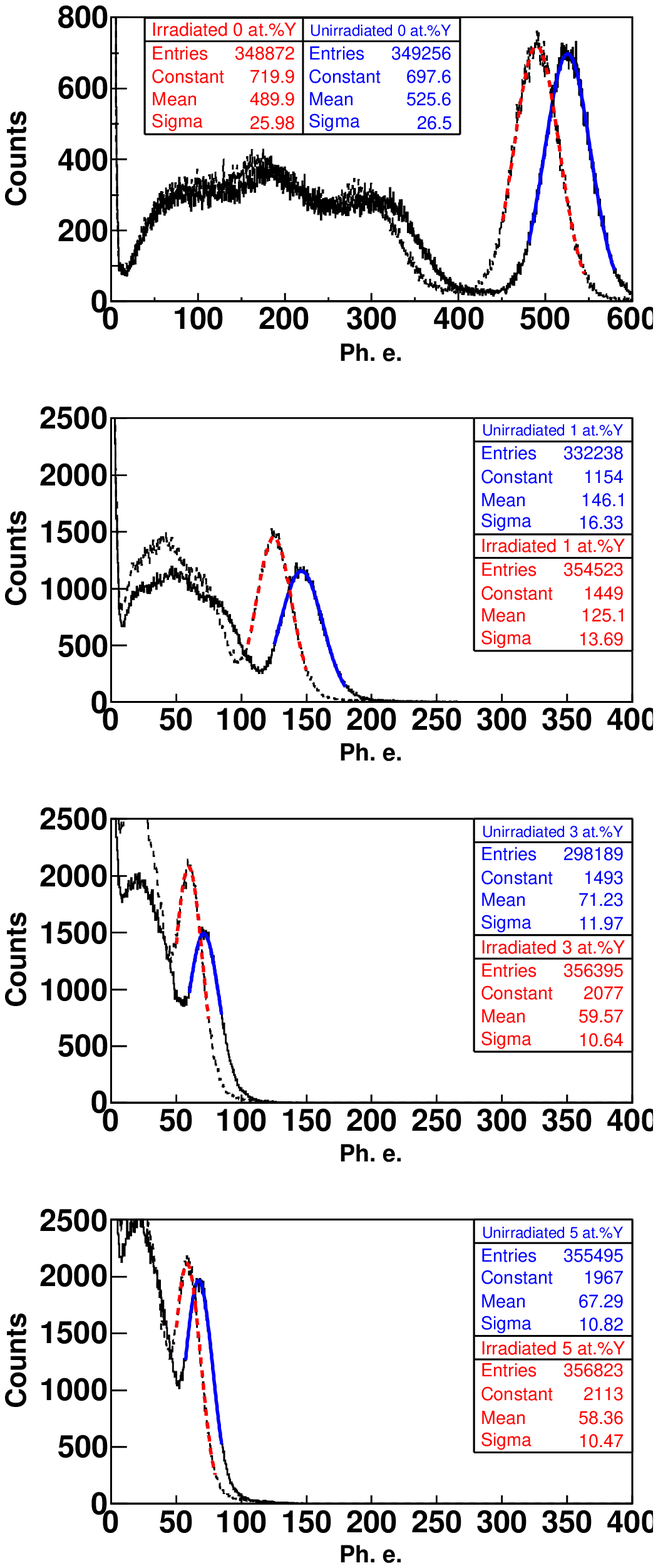}
     \includegraphics[width=0.32\textwidth]{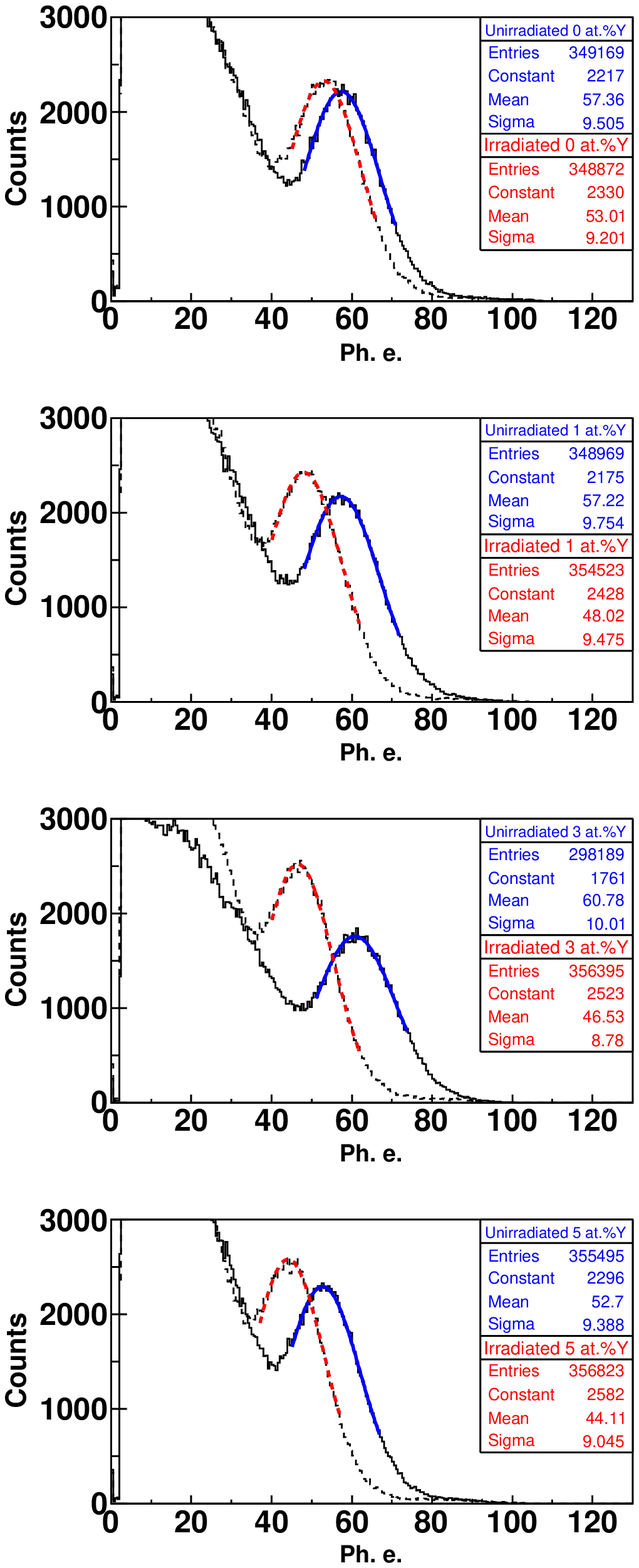}
  \end{center}
  \caption{Total (left column), slow (central column) and fast (right column) signals from the pure and yttrium doped BaF$_2$ samples before (solid blue lines) and after irradiation (dashed red lines). The yttrium fraction is 0, 1at.\%, 3at.\% and 5at.\% (from top to bottom).}
\label{fig:LO_Fast_Slow_Total}
\end{figure}

\begin{figure}[!ht]
  \begin{center}
    \includegraphics[width=0.75\textwidth]{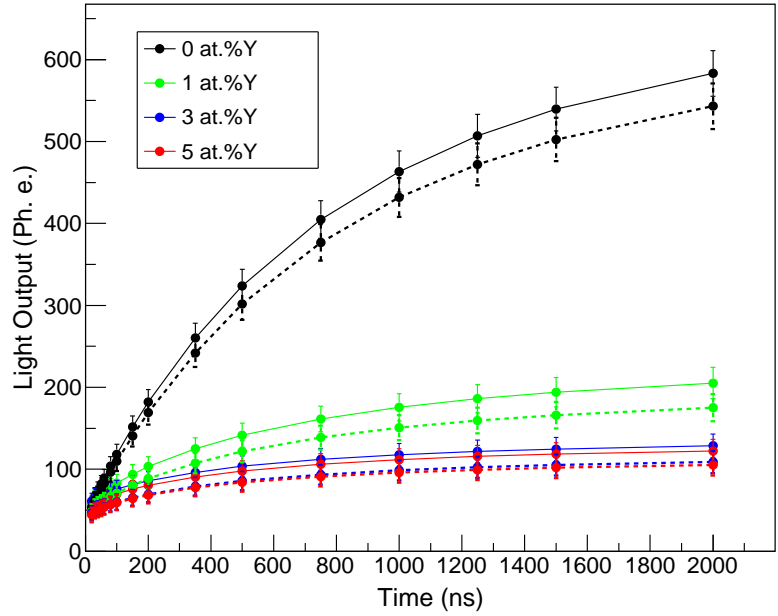}
  \end{center}
  %\vspace{3 cm}
  \caption{Light outputs of the pure and yttrium doped BaF$_2$ samples before (solid lines) and after (dashed lines) irradiation as a function of the integration time.}
\label{fig:kinetics}
\end{figure}

As mentioned above, the light output of the samples was estimated from the full absorption peaks of 511~keV gammas from the $^{22}$Na source. The spectra of the signals from the samples before and after neutron irradiation due to $^{22}$Na excitation are presented in fig.\ref{fig:LO_Fast_Slow_Total}.  The left column of fig.\ref{fig:LO_Fast_Slow_Total} shows the spectra of the total signals from the pure and yttrium doped BaF$_2$ samples while the central and right columns depict the spectra of the slow and fast luminescence components respectively. The spectra correspond, from top to bottom, to the samples with an yttrium fraction of 0\%, 1at.\%, 3at.\% and 5at.\% respectively. The solid blue lines show the spectra taken before irradiation, and the dashed red lines present the spectra of the irradiated samples. Figure~\ref{fig:kinetics} depicts kinetics of the signals from the same samples before (solid lines) and after (dashed lines) irradiation. Here the light outputs are shown as a function of the integration time.

As can be seen in fig.~\ref{fig:LO_Fast_Slow_Total} and fig.~\ref{fig:kinetics}, doping of the BaF$_2$ crystals with yttrium leads to decrease of the total signals. The main reason for the decrease in the total signal is the suppression of the slow emission component. In the  unirradiated samples the total signal drops by approximately a factor of 2.8, 4.5 and 4.8 in the 1at.\%, 3at.\% and 5at.\% yttrium doped samples respectively compared to the pure BaF$_2$ crystal. At the same time, the slow emission components in the same doped samples were suppressed by the 3.6, 7.4 and 7.8 respectively compared to the slow emission of the pure BaF$_2$ crystal. The fast emission in the sample doped with 1at.\%Y practically did not change in comparison with the fast emission of the pure BaF$_2$ crystal, while the fast emission components of the 3at.\% and 5at.\% yttrium doped samples decreased by about 2.5\% and 8\% respectively.

\begin{figure}[!ht]
  \begin{center}
    \includegraphics[width=0.75\textwidth]{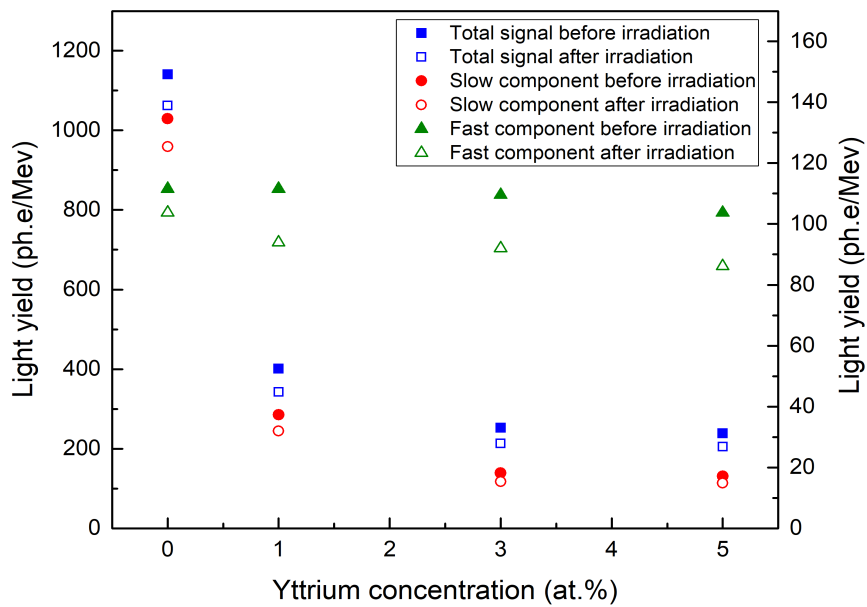}
  \end{center}
  %\vspace{3 cm}
  \caption{Light outputs of the total, slow and fast emissions of luminescence of the samples before and after irradiation, depending on the fraction of yttrium. The light outputs of the total and slow components are shown on the left scale while the light output of the fast component is on the right scale. }
\label{fig:LO_all}
\end{figure}

Light outputs of all samples (in ph.e/MeV) before and after irradiation are presented in fig.~\ref{fig:LO_all}. The fast and slow emission components are shown together with the total signals. Note that the total and slow emission light yields are shown on the left vertical scale while the fast emission component is on the right scale.

\begin{figure}[!ht]
  \begin{center}
     \includegraphics[width=0.7\textwidth]{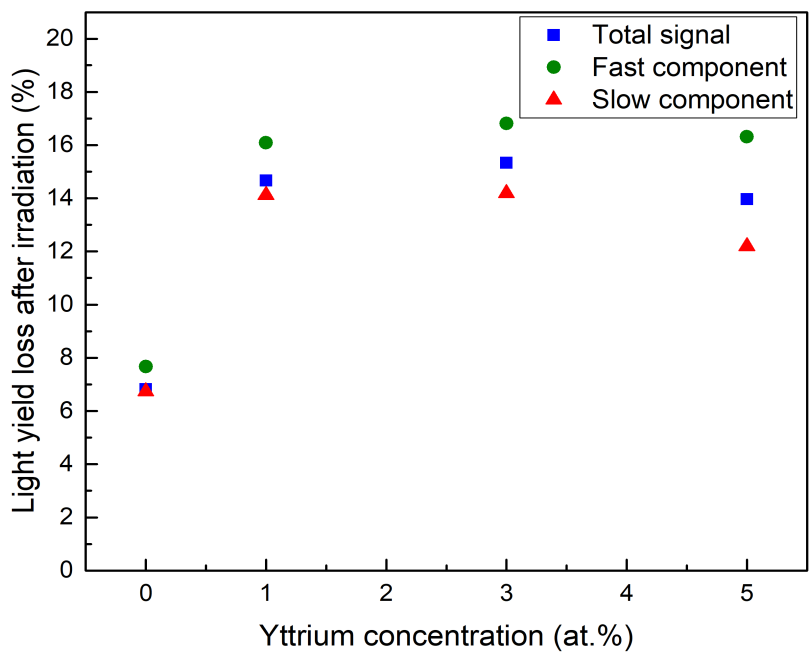}
  \end{center}
  %\vspace{3 cm}
  \caption{Light output loss of the pure and yttrium doped BaF$_2$ samples after irradiation depending on the fraction of yttrium.}
\label{fig:LO_loss}
\end{figure}

It is known that rare earth doped BaF$_2$ crystals usually lose their radiation hardness compared to the pure BaF$_2$. Our data confirms this. This can be seen qualitatively from the spectra in fig.\ref{fig:LO_Fast_Slow_Total} and from the light outputs of the samples before and after irradiation shown in fig.~\ref{fig:LO_all}. The light output loss of the total signals, fast or slow components of each sample was estimated as 1-LY$_{irr}$/LY$_0$, where LY$_0$ and LY$_{irr}$ are the light outputs (total, fast or slow component) of each sample before and after irradiation. The light output loss of each sample after irradiation for the total signals, slow and fast emissions is shown in fig.\ref{fig:LO_loss}. One can see that at our irradiation dose the light output loss of the pure BaF$_2$ crystal is about 7\%. The light output loss of the yttrium doped samples is approximately two times higher than that of the pure BaF$_2$ sample. Another interesting observation is that in all yttrium doped samples the light output loss of the fast emission component is 2-3\% higher than that of the slow emission.

Thus, our data confirm the suppression by several times of the slow luminescence component of yttrium doped BaF$_2$ crystals. However, the loss of the light output of the yttrium doped samples turned out to be two times higher than that of the pure BaF$_2$ crystal. Special attention should be paid to the fact that in the yttrium doped samples the loss of the light output of the fast luminescence component is 2-3\% higher than that of the slow component.

\section{Conclusion}
\label{sec:conclusion}

Light outputs of a pure BaF$_2$ crystal and three samples doped with yttrium in a proportion of 1at.\%, 3at.\% and 5at.\% were measured before and after irradiation in a neutron beam.
All samples were irradiated in the neutron beam of the IBR-2M pulsed reactor at JINR, Dubna. The total fast neutron (E>1~MeV) fluence that passed through the samples was about 2.3$\times$10$^{14}$~n/cm$^2$.

At this level of irradiation, the light output loss of the pure BaF$_2$ crystal is about 7\%, while the loss of the light output of the fast component is about 1\% greater than that of the slow one. The results show that the light output losses of both the fast and the slow emission component in the yttrium doped samples after irradiation are approximately two times higher than those in the pure BaF$_2$ crystal. It should be noted that the fast emission component of each yttrium doped sample after irradiation has a relative light output loss 2-3\% higher than the slow one. Undoubtedly, additional studies of light yield losses in a wider range of radiation doses are needed.

\acknowledgments

The authors are grateful to Dr. M.Bulavin (JINR) for help in irradiation of samples.

This work was supported by a grant from the Russian Foundation for Basic Research No.~18-52-05021.


\begin{thebibliography}{99}

\bibitem{Ershov} N.N.~Ershov et al., \emph{Studies in spectra and kinetics of intrinsic luminescence of fluorite-type crystals}, Optics and Spectroscopy, {\bf 53} (1982) p.51.
\bibitem{Gudovskikh} V.A.~Gudovskikh et al., \emph{Emission of Singlet and Triplet Excitons in Fluorite-Type Crystals under X-ray Excitation}, Optics and Spectroscopy, {\bf 53} (1982) p.910.
\bibitem{Laval} M. Laval et al., \emph{Barium fluoride inorganic scintillator for subnanosecond timing}, Nucl.Instr.Meth. in Physics Research {\bf 206} (1-2) (1983) p.168–176.
\bibitem{Mu2e_EMC} N.~Atanov et al., \emph{Design and Status of the Mu2e Crystal Calorimeter}, IEEE Trans.Nucl.Sci. {\bf 65}(2018) no.~8, p.2073.
\bibitem{Mu2e-II} F. Abusalma et al. (Mu2e collaboration), \emph{Expression of Interest for Evolution of the Mu2e Experiment},  arXiv:1802.02599.
\bibitem{Hitlin} D.~Hitlin et al. \emph{An APD for the Detection of the Fast Scintillation Component of BaF$_2$}, IEEE Trans.Nucl.Sci. {\bf 63}(2016) no.~2, p.513.
\bibitem{Artikov_filter} A.M.~Artikov et al., \emph{Suppression of the slow component of BaF$_2$ crystal luminescence with a thin multilayer filter}, J.Phys.Conf.Ser. {\bf 1162}(2019) 1, 012028. DOI: 10.1088/1742-6596/1162/1/012028
\bibitem{Atanov_cathode} N.~Atanov et al., \emph{A Photomultiplier With an AlGaN Photocathode and Microchannel Plates for BaF$_2$ Scintillator Detectors in Particle Physics}, IEEE Trans.Nucl.Sci. {\bf 67}(2020) no. 7, p.1760.
\bibitem{Chen} J.~Chen et al., \emph{Slow Scintillation Suppression in Yttrium
Doped BaF2 Crystals}, IEEE Trans.Nucl.Sci. {\bf 65}(2018), no. 8, p. 2147.
\bibitem{Schotanus} P.~Schotanus et al., \emph{Development study of a new gamma camera}, IEEE Trans.Nucl.Sci. {\bf NS-34}(1987) no. 1, p.272.
\bibitem{Woody} C.L.~Woody,P.W.~Levy and J.A.~Kierstead, \emph{Slow Component Suppression and Radiation Damage in Doped BaF$_2$ Crystals}, IEEE Trans.Nucl.Sci. {\bf 36}(1989) no. 1, p.536.

\bibitem{IBR-2M} http://flnph.jinr.ru/en/facilities/ibr-2
\bibitem{IBR_Ch3} M. Bulavin et al., \emph{Irradiation facility at the IBR-2M reactor for investigation of material radiation hardness}, Nucl.Instr.Meth. {\bf B 343}(2015), p.26.

\end{thebibliography}
\end{document}